# Neutrino oscillations in matter of varying density

Paul M. Fishbane*
*Physics Dept. and Institute for Nuclear and Particle Physics,*
*Univ. of Virginia, Charlottesville, VA 22904*

Peter Kaus**
*Aspen Center for Physics, Aspen, CO 81611*

We consider two-family neutrino oscillations in a medium of continuously-varying density as a limit of the process in a series of constant-density layers. We construct analytic expressions for the conversion amplitude at high energies within a medium with a density profile that is piecewise linear. We compare some cases to understand the type of effects that depend on the order of the material traversed by a neutrino beam.

## I. Transition amplitudes for structured matter.

The problem of neutrino oscillations in matter is of obvious importance, and it is interesting to see to what extent and in what manner analytic solution is possible. In this note we discuss a general approach to the problem which allows us to solve in a new way the case of neutrino passage through matter whose density varies linearly with distance.

Under the assumption that a two-channel approximation to neutrino mixing holds, the amplitude $A^j$ for passage of a neutrino beam of energy $E$ through a medium of constant electron density, whose properties (density $N_j$ and thickness $x_j$) we label by $j$, is a $2 \times 2$ matrix whose indices label flavors. It is given by the expression

$$A^j = \cos\phi_j + i\sin\phi_j \cos\left(2\theta_j\right)\sigma_z - i\sin\phi_j \sin\left(2\theta_j\right)\sigma_x \qquad (1.1)$$

where the work [1] of Mikheyev and Smirnov and of Wolfenstein [MSW] showed that the effect of the matter is summarized by

* e-mail address pmf2r@virginia.edu
** e-mail address pkaus@futureone.com

$$\Delta m^2 \to \Delta m_k^2 = \Delta m^2 \sqrt{\left(\cos 2q - x_k\right)^2 + \sin^2 2q}$$
$$j \to j_k = \frac{\Delta m_k^2}{4E} x_k \tag{1.2}$$
$$q \to q_k \ni \sin^2 2q_k = \frac{\sin^2 2q}{\left(\cos 2q - x_k\right)^2 + \sin^2 2q}$$

and

$$x_k \equiv \frac{2EV_k}{\Delta m^2} \text{ with } V_k = \sqrt{2} G_F N_k \,. \tag{1.3}$$

The mass parameter $\Delta m^2 = m_2^2 - m_1^2$ is positive. We recover the vacuum result, Cabibbo angle $q$, for $V = 0$.

Let us now consider the elements of the amplitude $A^{123..} = A^1 A^2 ... A^n$ for passage through a series of layers labeled sequentially from $n$ to 1. These can be found either by trace techniques or by the following direct technique: We extract a factor of $\cos f_j$ from each factor $A^j$. Then the amplitude takes the form

$$A^{12\cdots n} = \cos f_1 \cos f_2 \cdots \cos f_n \prod_{j=1}^{n} b_j \tag{1.4}$$

where

$$b_j = 1 + \vec{s} \cdot \vec{B}_j \tag{1.5}$$

and

$$\vec{B}_j = i \tan f_j \left(-\sin 2q_j, 0, \cos 2q_j\right) \tag{1.6}$$

To find the ordered product over the $b_j$, we require products of the form $\left(\vec{s} \cdot \vec{B}_1\right)\left(\vec{s} \cdot \vec{B}_2\right)\cdots\left(\vec{s} \cdot \vec{B}_m\right)$, and these can be found recursively using the $m = 2$ result

$$\left(\vec{s} \cdot \vec{B}_1\right)\left(\vec{s} \cdot \vec{B}_2\right) = \vec{B}_1 \cdot \vec{B}_2 + i\vec{s} \left(\vec{B}_1 \times \vec{B}_2\right).$$

(In the recursive development, it is helpful that the vector $B$ has no $y$-component.) The general form for our product is thereby found to be

$$\left(\vec{s} \cdot \vec{B}_1\right)\left(\vec{s} \cdot \vec{B}_2\right)\cdots\left(\vec{s} \cdot \vec{B}_m\right) = i^n \left(\prod_{j=1}^{m} \tan f_j\right) \times \begin{cases} \left[s_z \cos\left(2q_1 - 2q_2 + \cdots + 2q_m\right) - s_x \sin\left(2q_1 - 2q_2 + \cdots + 2q_m\right)\right] & m\,odd \\ \left[\cos\left(2q_1 - 2q_2 + \cdots - 2q_m\right) + is_y \sin\left(2q_1 - 2q_2 + \cdots + 2q_m\right)\right] & m\,even \end{cases} \tag{1.7}$$

In turn the transition matrix elements of interest are

$$A_{11}^{12\cdots n}=\left(\prod_{j=1}^{n}\cos f_j\right)\left\{\begin{array}{l}1+i\sum_{j=1}^{n}\tan f_j\,\cos 2q_j+(i)^2\sum_{j<k}\tan f_j\,\tan f_k\,\cos\left(2q_j-2q_k\right)\\ +(i)^3\sum_{j<k<\ell}\tan f_j\,\tan f_k\,\tan f_\ell\,\cos\left(2q_j-2q_k+2q_\ell\right)+\ldots\\ +(i)^n\tan f_1\tan f_2\cdots\tan f_n\,\cos\left(2q_1-2q_2+\cdots-(-1)^n\,2q_n\right)\end{array}\right\}\quad(1.8a)$$

and

$$A_{12}^{12\cdots n}=\left(\prod_{j=1}^{n}\cos f_j\right)\left\{\begin{array}{l}-i\sum_{j=1}^{n}\tan f_j\,\sin 2q_j+(-i)^2\sum_{j<k}\tan f_j\,\tan f_k\,\sin\left(2q_j-2q_k\right)\\ +(-i)^3\sum_{j<k<\ell}\tan f_j\,\tan f_k\,\tan f_\ell\,\sin\left(2q_j-2q_k+2q_\ell\right)+\ldots\\ +(-i)^n\tan f_1\tan f_2\cdots\tan f_n\,\sin\left(2q_1-2q_2+\cdots-(-1)^n\,2q_n\right)\end{array}\right\}\quad(1.8b)$$

One can immediately check that $|A_{11}|^2 + |A_{12}|^2 = 1$.

***Recovery of the vacuum result***. A useful check on Eqs. (1.8) follows from the assumption that the parameter $x_j$ is small compared to $\cos 2q$ for all $j$. We can thus approximate $q_j \cong q$ as well as $\Delta m_j^2 \cong \Delta m^2$. Assuming that the $n$ slabs have equal width, we in addition have $\tan f_j \cong \tan(f_{full}/n)$, where $f_{full} = \Delta m^2 X/(4E)$ is the angle appropriate to passage of a total length $X$ through vacuum. Thus when there are an odd number of angles,

$$2q_i - 2q_j + ... + 2q_k \cong 2q,$$

and when there are an even number of angles

$$2q_i - 2q_j + ... - 2q_k \cong 0.$$

Thus, with $f \equiv f_{full}/n$, the 11 element of the amplitude is

$$\begin{aligned}A_{11}&=\cos^n f\left\{\begin{array}{l}\left[1+\frac{n(n-1)}{2}(i\tan f)^2+\ldots\right]\\ +\cos 2q\left[n(i\tan f)+\frac{n(n-1)(n-2)}{3!}(i\tan f)^3+\ldots\right]\end{array}\right\}\\ &=\frac{1}{2}\left\{\left[e^{inf}+e^{-inf}\right]+\cos 2q\left[e^{inf}-e^{-inf}\right]\right\}\qquad(1.9a)\\ &=\cos f_{full}+i\cos 2q\sin f_{full}.\end{aligned}$$

We have here used

$$\frac{1}{2}\left[(1+x)^n-(1-x)^n\right]=nx+\frac{n(n-1)(n-2)}{3!}x^3+\ldots$$

$$\frac{1}{2}\left[(1+x)^n+(1-x)^n\right]=1+\frac{n(n-1)}{2!}x^2+\ldots$$

Similarly,

$$
\begin{aligned}
A_{12} &= \cos^n f \left\{ -\sin 2q \left[ n\left(i \tan f\right) + \frac{n(n-1)(n-2)}{3!}\left(i \tan f\right)^3 + \ldots \right]\right\} \\
&= \cos^n f \left\{ -\frac{1}{2}\sin 2q \left[ \left(1 + i \tan f\right)^n - \left(1 - i \tan f\right)^n \right]\right\} \\
&= -\frac{1}{2}\sin 2q \left[ e^{inf} + e^{-inf} \right] \\
&= -i \sin 2q \sin f_{full} .
\end{aligned}
\tag{1.9b}
$$

The expressions (1.9) match the vacuum result for a single width $X$, Eq. (1.1) with the angles replaced by their vacuum values.

***A Given ordering and its reverse***. One of the important features of oscillations within matter is that the amplitude for transitions depends on the order of the density of the layers through which a beam passes. The amplitude for passage through a single layer, Eq. (1.1), is symmetric, $A_j = A_j^T$. Therefore

$$A^{n\cdots 21} = \left(A^{12\cdots n}\right)^T , \tag{1.10}$$

where we recall that the sequence of superscripts matches the layer order. That means in particular that the diagonal (survival) elements are equal, e.g.,

$$A_{11}^{n\cdots 21} = A_{11}^{12\cdots n} \tag{1.11}$$

Unitarity in the two channel problem then gives us equal probabilities for the off-diagonal (conversion) elements for a given order and its reverse,

$$\left|A_{12}^{n\cdots 21}\right|^2 = \left|A_{12}^{12\cdots n}\right|^2 \tag{1.12}$$

This proof fails for the three-channel conversion problem [2].

To learn the relation between the off-diagonal elements of the amplitudes themselves, we note that $A^{12\ldots n}$ has the simple form

$$A^{12\cdots n} = \begin{bmatrix} a & b \\ -b* & a* \end{bmatrix}, \tag{1.13}$$

To see this, one can for example use a recursive proof: Equation (1.1) shows it is true for one layer, and explicit calculation shows that it is also true for two layers. Then one calculates

$$A^{12\cdots(n+1)} = A^{12\cdots n} A^{n+1} = \begin{bmatrix} a & b \\ -b* & a* \end{bmatrix} \begin{bmatrix} g & d \\ -d* & g* \end{bmatrix} = \begin{bmatrix} ag - bd* & ad + bg* \\ -b*g - a*d* & -b*d + a*g* \end{bmatrix},$$

and this has the requisite property.

We can now put Eqs. (1.10) and (1.13) together to show that

$$A^{n\cdots 21} = \begin{bmatrix} a & -b* \\ b & a* \end{bmatrix}. \tag{1.14}$$

Thus in particular

$$A_{12}^{n\cdots 21} = -\left(A_{12}^{12\cdots n}\right)* . \tag{1.15}$$

These results have been verified for the particular profiles we study below.

***High energy limit***. Let us call $x_{\min} = 2EV_{\min}/\Delta m^2$ the minimum value taken on by the parameter $x_j$ as $j$ runs from 1 to $n$. Then we can study a high energy limit,

$$x_{\min} >> \cos 2q,\ \sin 2q \tag{1.16}$$

In this limit

$$2q_j \cong \frac{\sin 2q}{x_j} \le \frac{\sin 2q}{x_{\min}} \ll 1 \tag{1.17}$$

and, because $\Delta m_j^2 \cong x_j\, \Delta m^2$ in this limit, we also have

$$f_j \cong \frac{1}{n} f_{full} x_j \tag{1.18}$$

For arbitrary $n$ this could be large or small; however, because we are ultimately interested in large $n$ and because for many situations this quantity is small in any case, we shall also assume, as part of the definition of the high energy limit, that $f_j$ is small for all $j$ and retain only first order terms in $f_j$. In particular the prefactor of the product over the $\cos f_j$ in Eqs. (1.9) is unity.

The effect of our limit is most easily seen in the 12 element of the transition amplitude $A$, Eq. (1.8b). Let us refer to the $m$-tuple sum in the curly brackets on the right of Eq. (1.8b) as $T_m$, so that $A_{12}$ is a sum over these sums.

Generally $T_m$ is a function of the potential $V$; however for $m = 1$, which is the one term in $A_{12}$ that is order independent, the potential dependence cancels,

$$T_1 = -i \sum_{j=1}^{n} \tan f_j \sin 2q_j = -i f_{full} \sin 2q \ . \tag{1.19}$$

This leading term essentially reproduces the vacuum result, independent of $n$. Both the potential and the order dependence are present in the $m = 2$ term,

$$\begin{aligned} T_2(V) &= -\sum_{j<k} \tan f_j \tan f_k \sin\left(2q_j - 2q_k\right) \\ &\cong -\left(\frac{1}{n} f_{full}\right)^2 \sin 2q \,\frac{2E}{\Delta m^2} \sum_{j<k}\left[V\left(x_k\right) - V\left(x_j\right)\right] \end{aligned} \tag{1.20}$$

This expression sets the pattern for the general term,

$$\begin{aligned} T_m(V) &= -\sum_{j_1<j_2<\ldots j_m} \tan f_{j_1} \cdots \tan f_{j_m} \sin\left(2q_{j_1} - 2q_{j_2} + \ldots - (-1)^m\, 2q_{j_m}\right) \\ &\cong \left(-i\frac{1}{n} f_{full}\right)^m \sin 2q \left(\frac{2E}{\Delta m^2}\right)^{m-1} S_m(V), \end{aligned} \tag{1.21}$$

where the multiple sum $S_m$ is defined by

$$S_m(V) = \sum_{j_1<j_2<\ldots j_m} V\left(z_1\right)\cdots V\left(z_m\right)\left[\frac{1}{V\left(z_1\right)} - \frac{1}{V\left(z_2\right)} + \ldots - (-1)^m \frac{1}{V\left(z_m\right)}\right]. \tag{1.22}$$

In the following section, we consider these sums for specific potentials.

The continuum limit of the sums of Eq. (1.22) can also be found in the usual manner. With the scaling variables $z_k \equiv j_k/n$, the large $n$ limit of $S_m(V)$ is

$$S_m(V) \cong n^m \int_0^1 dz_m \int_0^{z_m} dz_{m-1} \cdots \int_0^{z_2} dz_1 V\left(z_1\right)\cdots V\left(z_m\right)\left[\frac{1}{V\left(z_1\right)} - \frac{1}{V\left(z_2\right)} + \ldots - (-1)^m \frac{1}{V\left(z_m\right)}\right]$$

(1.23)

## II. Linear density profile

Here we consider a simple linear profile $V_1$ given by

$$V_1(x_j) = V_{\min} + V' jX/n. \tag{2.1}$$

The linear case has in fact been solved in other ways. In work [3, 4] on passage of neutrinos through layers of constant density matter, a linear density profile was used to interpolate the layers, and in so doing it was noticed that a formally identical problem had been solved much earlier in the context of atomic physics [5, 6]. This work was more thoroughly recalled and refined by Petcov [7]. More recently, another approach has produced a solution to the case of linear matter for an arbitrary number of channels [8]. In addition there is a body of work based on various approximations[9]. What we present here differs considerably in technique from the exact work cited and the high energy approximation that applies to the examples below complements the approximate work.

For the profile of Eq. (2.1) the double sum term, Eq. (1.22), takes the form

$$S_2(V_1) = V'\frac{X}{n}\sum_{j<k}(k-j) = V'\frac{X}{n}\sum_{k=1}^{n}\sum_{j=1}^{k-1}(k-j) = V'\frac{X}{n}\frac{n(n+1)(n-1)}{6} \tag{2.2}$$

This behaves as $n^2$ at large $n$, and since there is an additional factor $n^{-2}$ in $T_2$, Eq. (1.20), $T_2$ itself has a finite large $n$ limit, namely

$$\begin{aligned} T_2(V_1) &= -V'\frac{X}{n}\frac{n(n+1)(n-1)}{6}\left(\frac{1}{n}f_{full}\right)^2\left(\frac{2E}{\Delta m^2}\right)\sin 2q \\ &\xrightarrow[n\to\infty]{} -\frac{1}{12}V'X^2 f_{full}\sin 2q. \end{aligned} \tag{2.3}$$

The pattern is repeated for the general term[1]. We have

$$S_m(V_1) \cong \left(\frac{V'X}{n}\right)^{m-1}\sum_{j_1<j_2<\dots j_m}\left(j_2\cdots j_m - j_1 j_3\cdots j_m + \ldots - (-1)^m j_1\cdots j_{m-1}\right) \tag{2.4}$$

This multiple sum is explicitly calculable for any finite $n$ but is not very enlightening. The large $n$ behavior is simpler, and we give here the first results for the first few values of $m$:

$$\begin{aligned} S_2(V_1) &\xrightarrow[n\to\infty]{} (V'X)\, n^2\,(1/3!) \\ S_3(V_1) &\xrightarrow[n\to\infty]{} (V'X)^2\, n^3\,(7/5!) \\ S_4(V_1) &\xrightarrow[n\to\infty]{} (V'X)^3\, n^4\,(27/7!) \\ S_5(V_1) &\xrightarrow[n\to\infty]{} (V'X)^4\, n^5\,(321/9!) \\ S_6(V_1) &\xrightarrow[n\to\infty]{} (V'X)^5\, n^6\,(2265/11!) \\ S_7(V_1) &\xrightarrow[n\to\infty]{} (V'X)^6\, n^7\,(37575/13!) \end{aligned} \tag{2.5}$$

[1] We have ignored the terms containing $V_{\min}$ because they are nonleading as $n$ becomes large. In other words, the expression of Eq. (2.4) is already a large-$n$ approximation.

These results can be found either with the large $n$ behavior of the multiple sums of Eq. (1.22) or, more simply, with the multiple integrals of the large $n$ form Eq. (1.23).

We first remark that the factors of $n$ are those necessary to make the result finite, since $T_m$ contains an additional factor of $n^{-m}$. The sequence in the denominator can be identified [10] as follows:

$$T_m = \left(-i\right)^m \left(\frac{V'X^2}{2}\right)^{m-1} \frac{a\left(m-1\right)}{\left(2m-1\right)!} f_{full} \sin 2q = T_1 \left(-i\frac{V'X^2}{2}\right)^{m-1} \frac{a\left(m-1\right)}{\left(2m-1\right)!}, \qquad (2.6)$$

where $a(m)$ obeys the two-term recurrence relation

$$a(m+1) = a(m) + 2m(2m+1)a(m-1), \text{ with } a(0) = a(1) = 1. \qquad (2.7)$$

The quantity $a(m)$ is an expansion coefficient in several elementary functions, including some combinations of inverse trigonmetric functions and algebraic functions. Perhaps the most interesting relation is

$$\begin{aligned} \sum_{k=0}^{\infty} y^{2k} \frac{a(k)}{\left(2k+1\right)!} &= \frac{1}{y}\exp\left(y^2/2\right)\int_0^y e^{-u^2}\,du \\ &= \frac{1}{y}\exp\left(y^2/2\right)\frac{\sqrt{p}}{2}\operatorname{erf}\left(y\right) \\ &= \frac{1}{y}\exp\left(y^2/2\right)\sum_{k=0}^{\infty}\left(-1\right)^k \frac{y^{2k+1}}{\left(2k+1\right)k!}, \end{aligned}$$

a relation that can be applied to our case with the substitution

$$y = \exp\left(-ip/4\right)\sqrt{\frac{V'X^2}{2}}. \qquad (2.8)$$

The result is

$$A_{12}\left(V_1\right) = T_1 \frac{1}{y}\exp\left(y^2/2\right)\frac{\sqrt{p}}{2}\operatorname{erf}\left(y\right) \qquad (2.9)$$

## III. Comparison to other density profiles

It is instructive to compare the results of the previous section with two other density profiles of the same total thickness $X$, each representing a rearrangement of the matter that composes the profile $V_1$, i. e., each having the same integral of $V$ over $x$ from 0 to $X$. Specifically, we consider a profile peaked at the center,

$$V_2\left(x_j\right) = V_{\min} + \begin{cases} 2V'j\dfrac{X}{n} & j = 1,\ldots\dfrac{n}{2} \\ V'X - 2V'\left(j - \dfrac{n}{2}\right)\dfrac{X}{n} & j = \dfrac{n}{2}+1,\ldots n \end{cases} \qquad (3.1)$$

and a constant profile,

$$V_3(x_j) = V_{\min} + V'X/2. \qquad (3.2)$$

We again work at our high energy regime.

The profile $V_2$ can be treated by many of the same techniques that we used for $V_1$, even if the algebra is rather more complicated. We find for the double sum term, Eq. (1.22) with $m$ = 2,

$$S_2\left(V_2\right)=\sum_{j<k}\left[V_2\left(x_k\right)-V_2\left(x_j\right)\right]=-\frac{1}{2}V'Xn\,. \tag{3.3}$$

This result should be compared to the corresponding one for $S_2(V_1)$, Eq, (2.2), which is proportional to $n^2$ at large $n$. The term $T_2(V_2)$ vanishes in the large-$n$ limit. The cancellation is a consequence of the symmetry of the matter distribution about $X/2$ (as will be confirmed below for the $V_3$ case); indeed it is generally true that $S_m(V_2) = 0$ for even values of $m$ in the large-$n$ limit. For odd values of $m$, however, there is a nonzero limit. We have worked through the first few odd-$m$ expressions for $S_m(V_2)$ for finite $n$ . These are not simple, even in the large-$n$ limit, unless we set $V_{\min}$ to 0, which we do to make the expressions clear[2]. Then the large-$n$ results for all $m$ are

$$S_m\left(V_2\right)\cong\begin{cases}0 & m\ \text{even}\\ \dfrac{n^m\left(V'X\right)^{m-1}}{\left(2m-1\right)!!} & m\ \text{odd}\end{cases} \tag{3.4}$$

This gives for the conversion amplitude

$$A_{12}\left(V_2\right)=T_1\sum_{m=1,3\ldots}\left(-\frac{i}{2}V'X^2\right)^{m-1}\frac{1}{\left(2m-1\right)!!} \tag{3.5}$$

where $T_1$ is the amplitude for passage through a single layer of vacuum, Eq. (1.19). The sum on the right is a Lommel function [11] $U_{\frac{1}{2}}\left(w,0\right)=\sqrt{\frac{2w}{\pi}}\int_0^1\cos\left[\frac{1}{2}w\left(1-t^2\right)\right]dt$, and

$$A_{12}\left(V_2\right)=T_1\sqrt{\frac{\pi}{V'X^2}}\,U_{\frac{1}{2}}\left(\frac{1}{2}V'X^2,0\right). \tag{3.6}$$

The Lommel function is associated with diffraction from edges.

The amplitude for $V_3$ is simply given by the MSW result of Eq. (1.1), i.e.

$$A_{12}\left(V_3\right)=-i\sin2\theta_{V_3}\sin\phi_{V_3}\,.$$

In the "high energy" limit we have $\phi_{V_3}\cong\phi_{full}\xi_{V_3}=\frac{1}{2}V_3X$ and the conversion amplitude becomes

$$\begin{aligned}A_{12}\left(V_3\right)&\cong-i\frac{\sin2\theta}{\xi_{V_3}}\sin\left(\frac{1}{2}V_3X\right)=-i\frac{\phi_{full}\sin2\theta}{\phi_{full}\xi_{V_3}}\sin\left(\frac{1}{2}V_3X\right)\\&=T_1\frac{\sin\left(\frac{1}{2}V_3X\right)}{\phi_{full}\xi_{V_3}}=T_1\frac{\sin\left(\frac{1}{2}V_3X\right)}{\frac{1}{2}V_3X}.\end{aligned} \tag{3.7}$$

---

[2] Strictly speaking, this violates our high energy condition Eq. (1.16); however, because $V$ is only non-zero for an arbitrarily small range of $x$ this should not be troublesome.

The complete high energy expressions of Eqs. (3.6) and (3.7) are not very useful as grounds for comparison with the full result for $V_1$, Eq. (2.9). It is more transparent to expand each result for small values of $V'X^2$ (as well as for $V_{\min} = 0$). In that case,

$$A_{12}(V_1) \cong T_1\left\{1 - \frac{1}{3!}\frac{i}{2}V'X^2 - \frac{7}{5!}\left(\frac{1}{2}V'X^2\right)^2 + \ldots\right\} \tag{3.8a}$$

$$A_{12}(V_2) \cong T_1\left\{1 - \frac{1}{5!!}\left(\frac{1}{2}V'X^2\right)^2 + \ldots\right\} \tag{3.8b}$$

$$A_{12}(V_3) \cong T_1\left\{1 - \frac{1}{24}\left(\frac{1}{2}V'X^2\right)^2 + \ldots\right\} \tag{3.8c}$$

From these the conversion probabilities are, to leading order in $V'X^2$,

$$P_{12}(V_1) \cong |T_1|^2\left\{1 - \frac{4}{45}\left(\frac{1}{2}V'X^2\right)^2 + \ldots\right\} \tag{3.9a}$$

$$P_{12}(V_2) \cong |T_1|^2\left\{1 - \frac{2}{15}\left(\frac{1}{2}V'X^2\right)^2 + \ldots\right\} \tag{3.9b}$$

$$P_{12}(V_3) \cong |T_1|^2\left\{1 - \frac{1}{12}\left(\frac{1}{2}V'X^2\right)^2 + \ldots\right\} \tag{3.9c}$$

The ratios of these probabilities is in principle subject to experimental test, although it is clear that such tests would be very difficult.

## Acknowledgements

We would like to thank the Aspen Center for Physics, where much of this work was done. PMF would also like to thank Dominique Schiff and the members of the LPTHE at Université de Paris-Sud for their hospitality. This work is supported in part by the U.S. Department of Energy under grant number DE-FG02-97ER41027.